\documentclass[aps,prl,reprint,twocolumn,superscriptaddress,floatfix,nofootinbib]{revtex4-1}
\usepackage{amsmath,amssymb,color,comment,physics}
\usepackage[makeroom]{cancel}
\usepackage[caption=false]{subfig}
\usepackage{amsfonts}
\usepackage{mathrsfs}
\usepackage{graphicx}
\usepackage{float}
\usepackage[countmax]{subfloat}
\usepackage{braket}
\usepackage{bm}
\usepackage[normalem]{ulem}
\usepackage[english]{babel}
\usepackage{dsfont}
\usepackage{umoline}
\usepackage[bookmarks=true,colorlinks,linkcolor=OrangeRed,urlcolor=NavyBlue,citecolor=RoyalBlue]{hyperref}
\usepackage[dvipsnames]{xcolor}

\usepackage{mathtools}
\newtagform{supplementary}[S.]()
\newtagform{main}[]()

\makeatletter
\def\maketitle{
	\@author@finish
	\title@column\titleblock@produce
	\suppressfloats[t]}
\makeatother
\begin{document}
	\title{Towards the continuum limit of a $(1+1)$d quantum link Schwinger model}

	\author{Torsten V. Zache}
	\affiliation{Center for Quantum Physics, University of Innsbruck, 6020 Innsbruck, Austria}
	\affiliation{Institute for Quantum Optics and Quantum Information of the Austrian Academy of Sciences, 6020 Innsbruck, Austria}
	\affiliation{Heidelberg University, Institut f{\"u}r Theoretische Physik, Philosophenweg 16, 69120 Heidelberg, Germany}
	\author{Maarten Van Damme}
	\affiliation{Department of Physics and Astronomy, University of Ghent, Krijgslaan 281, 9000 Gent, Belgium}
	\author{Jad C. Halimeh}
	\affiliation{INO-CNR BEC Center and Department of Physics,
		University of Trento, Via Sommarive 14, I-38123 Trento, Italy}
	\author{Philipp Hauke}
	\affiliation{INO-CNR BEC Center and Department of Physics,
		University of Trento, Via Sommarive 14, I-38123 Trento, Italy}
	\author{Debasish Banerjee}
	\affiliation{Saha Institute of Nuclear Physics, HBNI, 1/AF Bidhannagar, Kolkata 700064, India}
	\affiliation{Institut f\"ur Physik, Humboldt-Universit\"at zu Berlin, Zum Gro\ss en Windkanal 6, 12489 Berlin, Germany}
	\date{\today}
	
	\begin{abstract}
		The solution of gauge theories is one of the most promising applications of quantum technologies. Here, we discuss the approach to the continuum limit for $U(1)$ gauge theories regularized via finite-dimensional Hilbert spaces of quantum spin-$S$ operators, known as quantum link models. For quantum electrodynamics (QED) in one spatial dimension, we numerically demonstrate the continuum limit by extrapolating the ground state energy, the scalar, and the vector meson masses to large spin lengths $S$, large volume $N$, and vanishing lattice spacing $a$. By exactly solving Gauss' law for arbitrary $S$, we obtain a generalized PXP spin model and count the physical Hilbert space dimension analytically. This allows us to quantify the required resources for reliable extrapolations to the continuum limit on quantum devices. We use a functional integral approach to relate the model with large values of half-integer spins to the physics at topological angle $\Theta=\pi$. Our findings indicate that quantum devices will in the foreseeable future be able to quantitatively probe the QED regime with quantum link models.
	\end{abstract}
	
	\maketitle
	
	\paragraph{Introduction --}
	The rapid development of quantum technologies culminating in the precise control of large quantum systems~\cite{BlochDalibard2012, BlattRoos2012,georgescu2014quantum,Hauke_2012} has fundamentally altered the scope of physics questions that can be addressed for strongly interacting systems. As a complementary approach to smashing nuclei in colliders to uncover their substructure, one can realize quantum many-body systems in controlled analog quantum simulators or digital quantum computers, and study their ground state properties at finite density or during real-time evolution associated with quenches, which are extremely difficult to tackle using Markov Chain Monte Carlo methods~\cite{Hofstetter_2018, Wiese_2014}.Motivated by this possibility, pioneering proposals~\cite{zohar2011confinement,Zohar:2012ay,Tagliacozzo:2012vg,Banerjee2012pg} have been put forward to study properties of lattice gauge theories with the long-term goal of simulating quantum chromodynamics (QCD), the theory of strong interactions. These ideas have triggered an intensive theory effort to devise efficient and feasible implementations (for recent reviews see, e.g.,~\cite{Zohar_2015, Dalmonte2016, banuls2020simulating}), resulting in first experimental realizations~\cite{martinez2016real, klco2018quantum, schweizer2019floquet, gorg2019realization, mil2020scalable,yang2020observation,atas20212, Lu_2019} in recent years.
	\begin{figure}[h]
		\centering
		\includegraphics[scale=0.42]{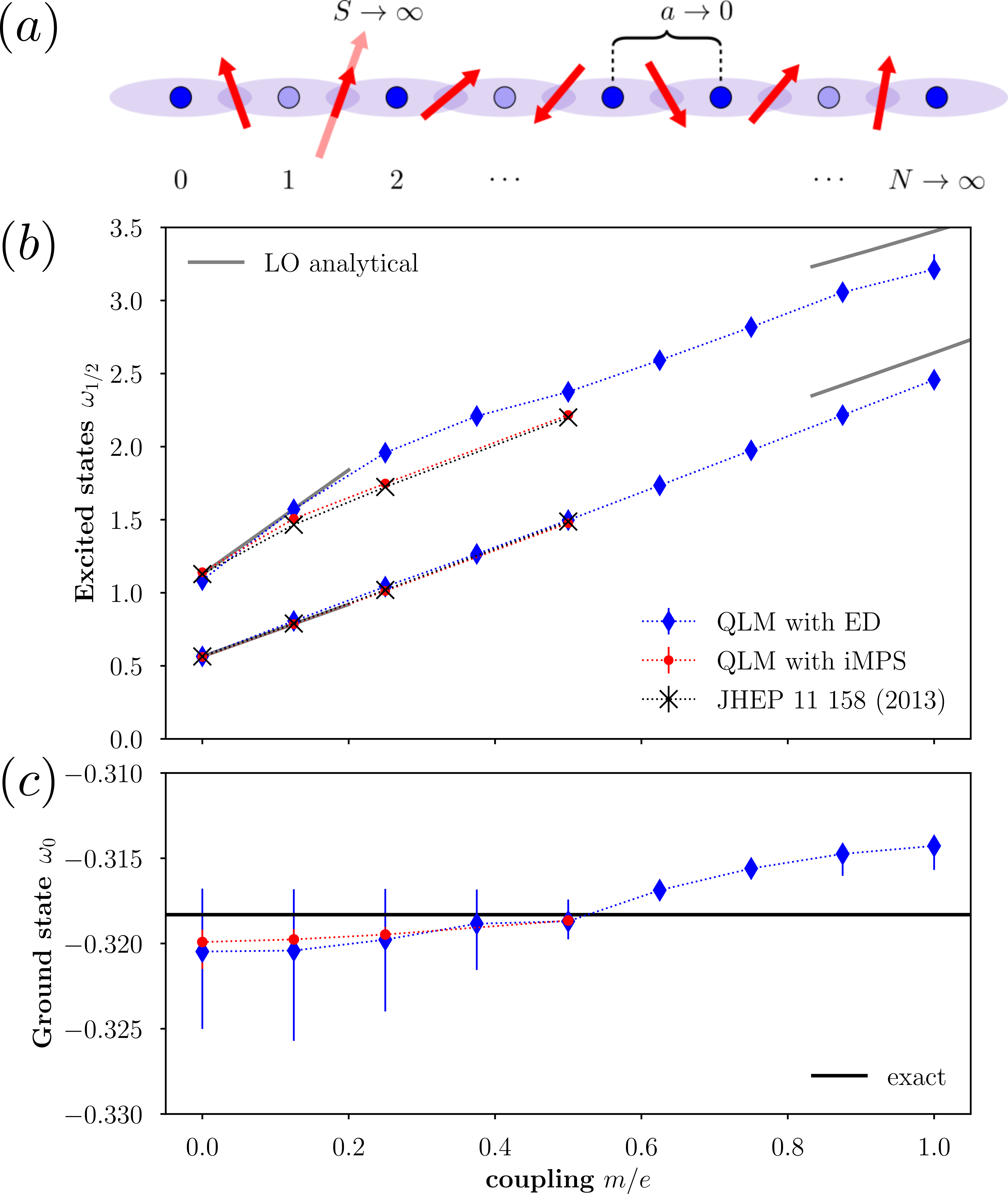}
		\caption{$\textbf{(a)}$ In a U(1) QLM, matter fields (blue dots) reside on the sites of a lattice while gauge fields, represented by spins (red arrows), live on the links connecting two neighboring lattice sites. Gauss' law ties consecutive gauge fields to the matter in between as indicated by the shaded ellipses.
			The ground state energy, shown in $\textbf{(c)}$, and the first two excited states (with vector and scalar quantum numbers), shown in $\textbf{(b)}$, in the zero momentum sector of QED obtained from the $U(1)$ QLM using ED and iMPS show excellent agreement with the analytical prediction for $m/e=0$. In $\textbf{(b)}$, the grey solid lines indicate the leading order (LO) analytical expansions~\cite{sriganesh2000new} for small $m/e$ and $e/m$, respectively. The error bars indicate an estimated systematic uncertainty (see Section IV in SM).}
		\label{fig:illustration}
	\end{figure}
	
	Nonperturbative calculations of quantum field theories (QFTs) require a careful treatment of regularization and renormalization, for which the lattice approach has proven most successful~\cite{montvay1997quantum}. The lattice Hamiltonian of a gauge theory~\cite{kogut1975hamiltonian} retains exact gauge invariance while a finite spatial lattice reduces the infinite number of degrees of freedom of the field theory to a finite number of lattice sites and links. The local Hilbert space dimension of the gauge fields, however, remains infinite in the original Wilsonian formulation. Quantum link models (QLMs)~\cite{Horn:1981kk, Orland:1989st, chandrasekharan1997quantum} regulate these infinite dimensional Hilbert spaces with qudits while maintaining exact gauge invariance. They are ideal candidates to be studied on quantum devices such as analog quantum simulators or digital quantum computers, which typically work with finite dimensional local Hilbert spaces. Extracting information relevant for the QFT in the continuum limit, especially when realized in the low-dimensional Hilbert spaces available in current quantum devices, requires a sequence of extrapolations which is the main topic of this article.
	
	Effects of truncating the infinite-dimensional Hilbert space have been extensively investigated for lattice theories with continuous global and local symmetries~\cite{Byrnes:2002nv, Byrnes:2005qx, Yang2016, Buyens_2017, Niedermayer:2016hzw, Raychowdhury:2018osk, davoudi2021toward, Yockey2018, Zhang2021, Gustafson2021}. For pure $U(1)$ gauge theory in $(2+1)-$d dimensions, truncations in the magnetic basis have been argued to be superior to truncations in the electric basis for reaching the continuum limit~\cite{Paulson:2020zjd} (see also~\cite{Kaplan:2018vnj,Yockey2019,bender2020gauge}). Nevertheless, rapid convergence in such truncations has been observed previously in~\cite{kuhn2014quantum,buyens2017finite}. The related approach of approximating continuous groups with discrete groups of increasing order also has a long history~\cite{Bhanot:1981xp, Hasenfratz:2000sa, Ercolessi:2017jbi, Alexandru:2019nsa, Kurkcuoglu2021, Alam2021, Bender2018, Hackett2019}.
	
	The continuum limit physics is quite sensitive to the nature of the employed truncation. 
	For example, in the $(1+1)-$d $O(3)$ model 
	the physics of asymptotic freedom could only be recovered with at least a 16-dimensional local Hilbert space in the angular momentum basis truncation~\cite{Bruckmann:2018usp}.
	However, using qubit operators~\cite{Singh:2019uwd}, it was shown that the same continuum limit only required 2-qubits per site~\cite{Bhattacharya:2020gpm}. The QLM approach is similar to this qubit-regularization, but uses larger spin-$S$ operators for the $U(1)$ gauge links. It was analytically shown that large representations for the gauge links in QLMs recover the standard Wilson lattice gauge theory ~\cite{Schlittgen:2000xg}. A fine-tuning free approach to the continuum limit using the QLMs is via the dimensional reduction in the D-theory formulation~\cite{Brower:2003vy}. This is, however, only possible if a phase with an exponentially large correlation length is generated~\cite{Beard:2004jr}. In this article, we explore the former approach and show that careful analysis techniques allow us to reach the continuum limit quantitatively for certain physical observables, even with very small values of $S$ ($\lesssim 3-5$), extending the observations made in~\cite{kuhn2014quantum,buyens2017finite}. We solve the Gauss law analytically for a general spin-$S$ representation, and estimate the quantum resources to simulate the continuum limit.
	Our results may help to improve digital quantum simulations of the massive Schwinger model~\cite{shaw2020quantum,chakraborty2020digital,honda2021digital}, through variational quantum eigensolvers and Trotterized real-time dynamics.
	Finally, we use path integrals to demonstrate how large half-integer spins give rise to the topological angle $\Theta=\pi$.

	\paragraph{Hamiltonian and Gauss' law --}
	We focus on $U(1)$ gauge theories in $d$ spatial dimensions with (staggered) fermionic matter~\cite{kogut1975hamiltonian}. The gauge fields are described by $\hat{\mathbf{S}}_{\mathbf{n},j}= (\hat{S}^+,\hat{S}^-,\hat{S}^z)_{\mathbf{n},j}$, the raising, lowering and z-component quantum spin-$S$ operators on the links $(\mathbf{n},j)$ connecting neighboring sites $\mathbf{n}$ and $\mathbf{n} + \mathbf{e}_j$ on a hypercubic lattice of size $N^d$~\footnote{Throughout this work, we employ periodic boundary conditions.}. For any $S$, the raising (lowering) operators are $S^\pm = (S^x \pm i S^y)/2$ with $S^{x/y}$ the $x/y$ component of the spin vector $\vec{S}$. For the spin-$1/2$ representation, for example, the spin operators are related to Pauli matrices, $\vec{S} = \vec{\sigma}/2$. The spins are coupled to fermionic operators $\hat{\psi}_{\mathbf{n}}$ on the sites, as described by the Hamiltonian
	\begin{align}\label{eq:LGT_Hamiltonian}
		\hat{H} &= \frac{g^2}{2} \sum_{\mathbf{n},j} \big(\hat{S}^z_{\mathbf{n},j}\big)^2 + \mu \sum_{\mathbf{n}} (-1)^{n_1 + \dots + n_d} \hat{\psi}_\mathbf{n}^\dagger\hat{\psi}_\mathbf{n} - \hat{H}_{\rm m} \nonumber \\
		&- \frac{1}{2\sqrt{S (S + 1)}}  \sum_{\mathbf{n},j} (-1)^{\sum_{k<j} n_k} \big( \hat{\psi}_\mathbf{n}^\dagger \hat{S}^+_{\mathbf{n},j}\hat{\psi}_{\mathbf{n}+\mathbf{e}_j} + \text{H.c.}\big) \;. 
	\end{align}
	The first two terms are the electric field energy at the bare coupling $g$, and the staggered fermion mass $\mu$.
	The gauge-matter interaction is the correlated hopping of fermions along a link with the simultaneous raising or lowering of the corresponding spin. For $d > 1$, there is the magnetic energy term, $\hat{H}_{\rm m} =\frac{1}{2g^2 S^2 (S + 1)^2} \sum_{P}\big(\hat{S}^+_{P_1} \hat{S}^+_{P_2}\hat{S}^-_{P_3} \hat{S}^-_{P_4}+ \text{h.c.}\big)$, with $P$ labeling elementary plaquettes consisting of links $P_{1,2,3,4}$ forming a square. 
	We identify the spins with gauge fields $(\hat{U}, \hat{U}^\dagger) \leftrightarrow (\hat{S}^+, \hat{S}^-)/\sqrt{S(S+1)}$ and electric field $\hat{E} \leftrightarrow \hat{S}^z$.
	This identification preserves the commutation relations, $[E,U^{(\dagger)}] =(-) U^{(\dagger)}$, 
	as well as an exact gauge symmetry with $(2S+1)$-dimensional Hilbert space. To achieve the correct scaling behavior, appropriate factors of $S$ are inserted in the dimensionless couplings in  Eq.~\eqref{eq:LGT_Hamiltonian} (see Sec I in SM for more details). The gauge transformations are generated by the Gauss law operator,
	\begin{align}  \label{eq:Glaw}
		\hat{G}_\mathbf{n} =  \sum_j \big(\hat{S}^z_{\mathbf{n}+\mathbf{e}_j,j} - \hat{S}^z_{\mathbf{n},j}\big) - \bigg[ \frac{ (-1)^\mathbf{n} - 1 }{2} +  \hat{\psi}_\mathbf{n}^\dagger\hat{\psi}_\mathbf{n} \bigg] \;,
	\end{align}
	satisfying $\big[\hat{H},\hat{G}_\mathbf{n}\big] = 0$. The Hilbert space thus separates into superselection sectors labelled by eigenvalues of $\hat{G}_\mathbf{n}$. For the physical Hilbert space, $\mathcal{H}_{\rm phys}$, we require $\hat{G}_\mathbf{n}|\text{phys} \rangle = 0$.
	
	In the renormalization group (RG) sense, the parameters $g$, $\mu$, and $S$ can be regarded as directions in the space of couplings, to be adjusted such that the theory flows to a fixed point corresponding to the desired QFT. In the remainder of this letter, we focus on the case of one spatial dimension, $d=1$, where Eq.~\eqref{eq:LGT_Hamiltonian} provides a lattice version of the (massive) Schwinger model~\cite{Schwinger1962, Coleman:1975pw, Coleman:1976uz}. The continuum Schwinger model is parameterized by the (bare) values of the electric charge $e$ and the fermion mass $m$. In its lattice version at lattice spacing $a$, these parameters appear through the dimensionless combinations $g=ae$ and $\mu=am$. The QFT limit is reached for large $S$, large $N$, and small $a$, as demonstrated in Fig.~\ref{fig:illustration}. Specifically, we first take the infinite spin length limit $S\rightarrow \infty$ at fixed $a$ and $N$; then the thermodynamic limit $N \rightarrow \infty$ at fixed $a$; and finally the continuum limit $a \rightarrow 0$ at fixed $\mu/g$. The different extrapolations have to be performed for appropriately rescaled (``renormalized'') quantities that correspond to physical observables (see Sec IV in the SM for details). In general, the final continuum limit involves a rescaling of the dimensionless coupling constants in order to reach the RG fixed point~\cite{montvay1997quantum}. This complication is absent for the present model (except for a redefinition of the ground state energy)~\cite{Coleman:1975pw,abdalla1991non}.

	\begin{figure*}
		\centering
		\includegraphics[width=0.66\columnwidth]{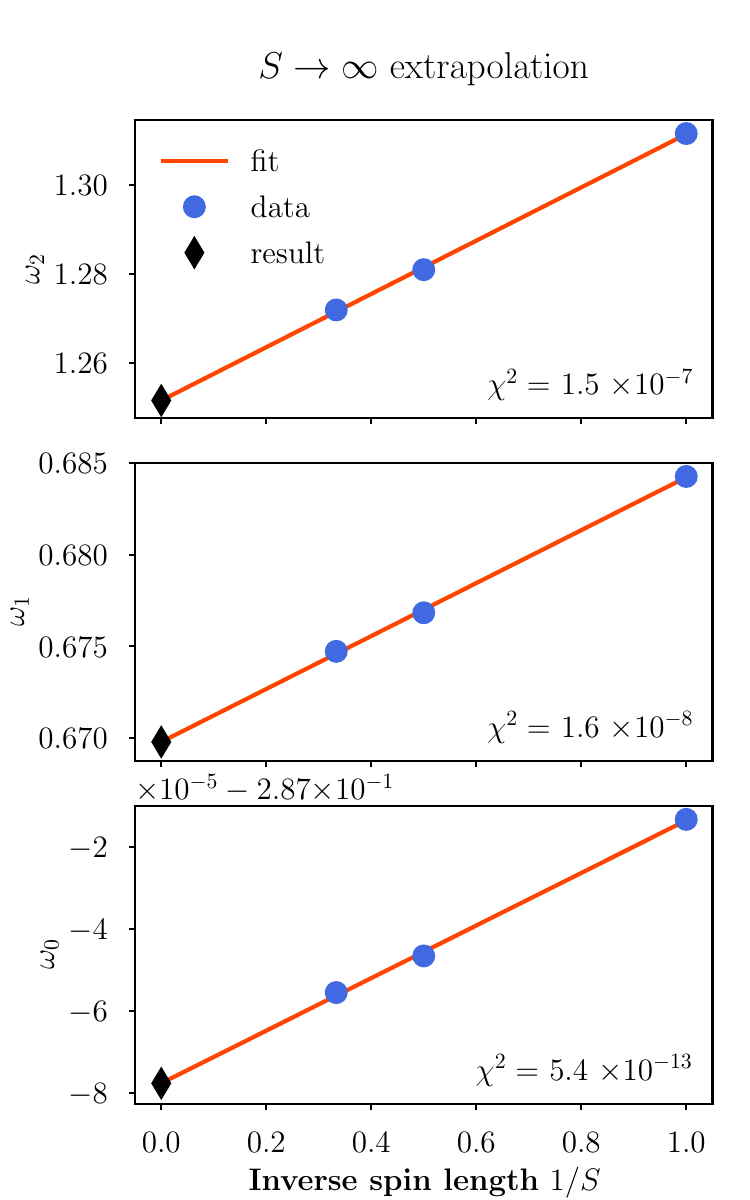}
		\includegraphics[width=0.66\columnwidth]{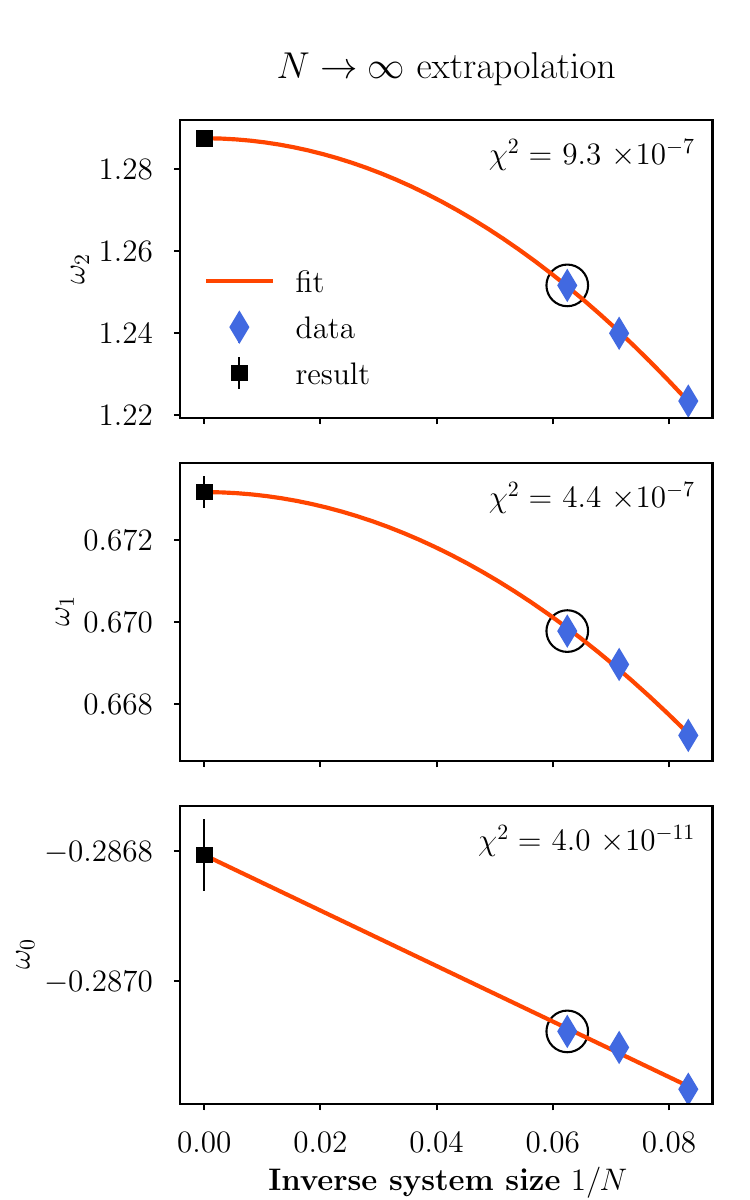}
		\includegraphics[width=0.66\columnwidth]{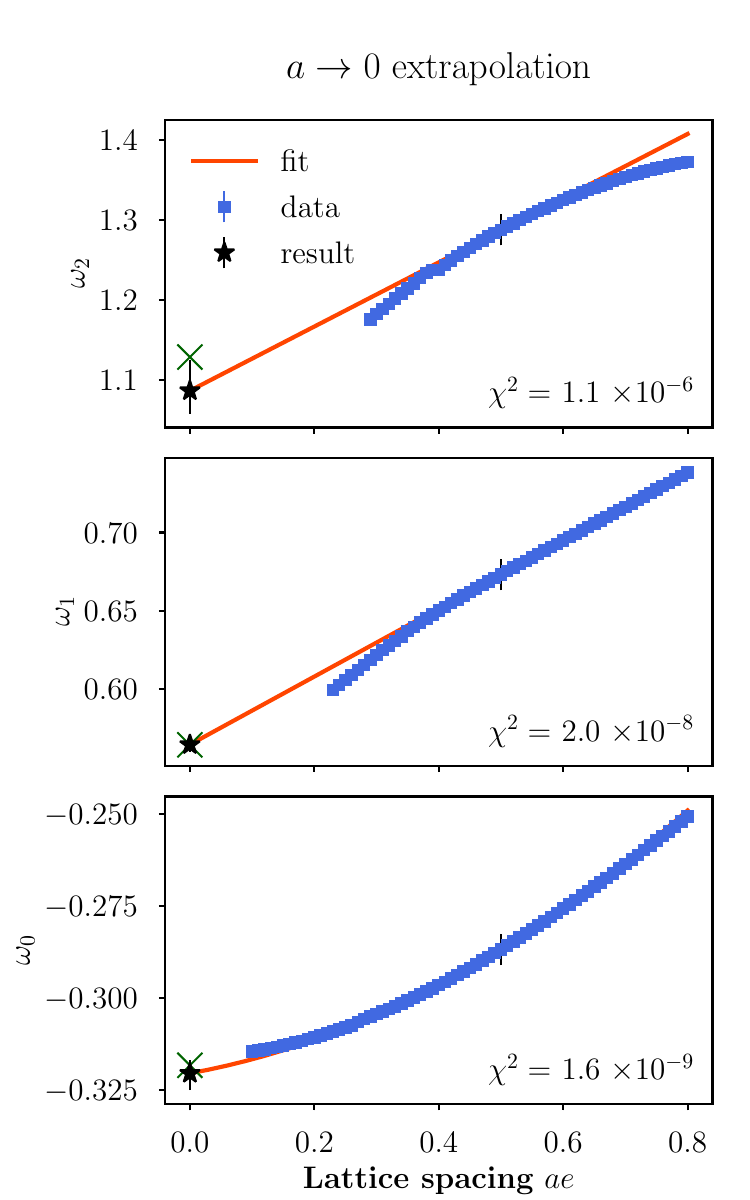}
		\caption{We illustrate the sequence of extrapolations in the panels from the left to the right, required to reach the continuum limit for the ED data with $m/e = 0$. The energies of the vacuum (bottom row), vector particle (middle row), and scalar particle (top row) are extrapolated to $S \rightarrow \infty$ (left column), $N\rightarrow \infty$ (middle column), and $a\rightarrow 0$ (right column), as discussed in the main text.
			The circles in the middle row indicate the values obtained from the corresponding $S$-extrapolations shown in the left column. Similarly, the ticks in the right column indicate the values corresponding to the $N$-extrapolations in the middle column. For clarity, we only show selected values of $ae$ and $N$ for the first two extrapolations. For comparison, the green crosses indicate the exact analytical results. All fits employed for the extrapolations are polynomials { and $\chi^2$ denotes the resulting normalized square error of the fit} (see Sec IV of the SM for details, where we also provide a table with our quantitative results).}
		\label{fig:extrapolations}
	\end{figure*}
	
	\paragraph{Mass spectrum from the QLM --}
	The massive Schwinger model is considerably simpler than QED in higher dimensions due to the absence of magnetic interactions and the strong Gauss' law constraints. Consequently, both the weak and strong coupling limits, $e/m = 0$ and $e/m=\infty$, respectively, are exactly solvable, and analytic expansions around these limits~\cite{Hamer:1997dx,sriganesh2000new} can be used for benchmarking the extrapolation. Our principal numerical methods are exact diagonlization (ED) (using the Python package QuSpin~\cite{weinberg2019quspin}) and variational techniques based on infinite Matrix Product States (iMPS)~\cite{zauner2018variational,haegeman2012variational} directly in the thermodynamic limit. To perform ED, we derive an equivalent spin model constrained by a projector $\mathcal{P}$ on neighboring gauge link configurations allowed by Gauss' law. The resulting Hamiltonian has the form (see Sec II in the SM):
	\begin{align}\label{eq:spin_model}
		\hat{H} = \mathcal{P} \sum_{n=1}^{N} \bigg\lbrace \frac{g^2}{2} \big(\hat{S}^z_{n}\big)^2 +  2\mu(-1)^n  \hat{S}^z_{n}   - \frac{\hat{S}^x_{n} }{\sqrt{S(S+1)}}  \bigg\rbrace \mathcal{P} \;.
	\end{align}
	For $S=1/2$, this reduces to a constrained model of hard-core bosons~\cite{Fendley2004}, sometimes referred to as PXP model, whose total number of allowed states can be analytically counted~\cite{Turner2018}, and which is believed to explain the anomalous thermalization observed in the 51-Rydberg atom experiment~\cite{bernien2017probing}. The relation between a PXP model and the spin-1/2 QLM was first noted in \cite{surace2020lattice}, which we generalize here to arbitrary $S$. Moreover, we extend the analysis of \cite{Turner2018}, and derive an analytic expression for the dimension of the physical Hilbert subspace, given by (see SM)
	\begin{align}\label{eq:H_dim}
		\text{dim} \, \mathcal{H}_\text{phys} (S,N) = 2^N \sum_{m=1}^{2S+1} \left[\cos \left(\frac{m\pi}{4S + 3}\right)\right]^N \;.
	\end{align}
	The remarkably small Hilbert space size, scaling linearly with $S$ at a fixed $N$ (see Sec. II in the SM for an illustration), enables ED calculations for relatively large system sizes, and data up to $S=3$ and $N=16$ is presented here. Results from iMPS simulations in the thermodynamic limit are also shown for $S \le 5$ for the Hamiltonian in Eq.~\eqref{eq:LGT_Hamiltonian}, where Gauss' law is enforced by adding a large energy penalty $\propto \sum_\mathbf{n}\hat{G}^2_\mathbf{n}$~\cite{Halimeh2020, vandamme2020gaugesymmetry}, with $\hat{G}_\mathbf{n}$ defined in Eq.~\eqref{eq:Glaw}. Our results, summarized in Fig.~\ref{fig:illustration}, demonstrate how to accurately reach the continuum limit with QLMs for the ground state energy and the energies of the first two excited states, the ``vector'' and ``scalar'' particles. We find excellent agreement in the strong coupling limit, due to small fluctuations around $\langle\hat{S}^z\rangle = 0$, such that our numerics with small $S$ ($\le 3$ for ED and $\le 5$ for iMPS) already capture the relevant physics. For higher excited states or towards weak coupling, the fluctuations grow more pronounced and larger spin lengths $S$ become necessary~\footnote{The path integral construction presented in the Sec V of the SM suggets that the U($1$) limit can be reached by scaling the spin length as $S \propto 1/g$ for a fixed, but large $gS \gg 1$.}.

	The detailed steps of the underlying extrapolation are shown in Fig.~\ref{fig:extrapolations} for ED, illustrated for the analytically solvable strong-coupling limit ($m/e=0$). Despite small spin lengths $S=1,2,3$, we observe a clear $1/S$ scaling, enabling a reliable extrapolation to the $S\rightarrow \infty$ limit. Similarly, the subsequent $N\rightarrow \infty$ extrapolation is performed with the expected leading behavior at large $N$. The largest systematic error arises from the choice of fit range for the final $a \rightarrow 0$ extrapolation, which require increasingly large values of $N$ and $S$, attributed to increasing electric field fluctuations at the continuum limit. Empirically, we find that systematic errors are minimized by disregarding ``far-off'' $N$ and $S$ extrapolation where the extrapolated values differ by more than $10 \%$ from the one of the largest available system size. We thus select a smallest lattice spacing $a$ for which the underlying data is sufficiently converged with respect to $S$ and $N$. Note that this procedure naturally depends on the observable. The ground state energy can be extrapolated with lattice spacings down to $ae\sim0.1$, while we only reach $ae\sim0.3$ for the scalar mass. Details of the numerical extrapolations are presented in Sec IV of the SM.
	
\begin{figure}[ht]
		\centering
		\includegraphics[scale=0.4]{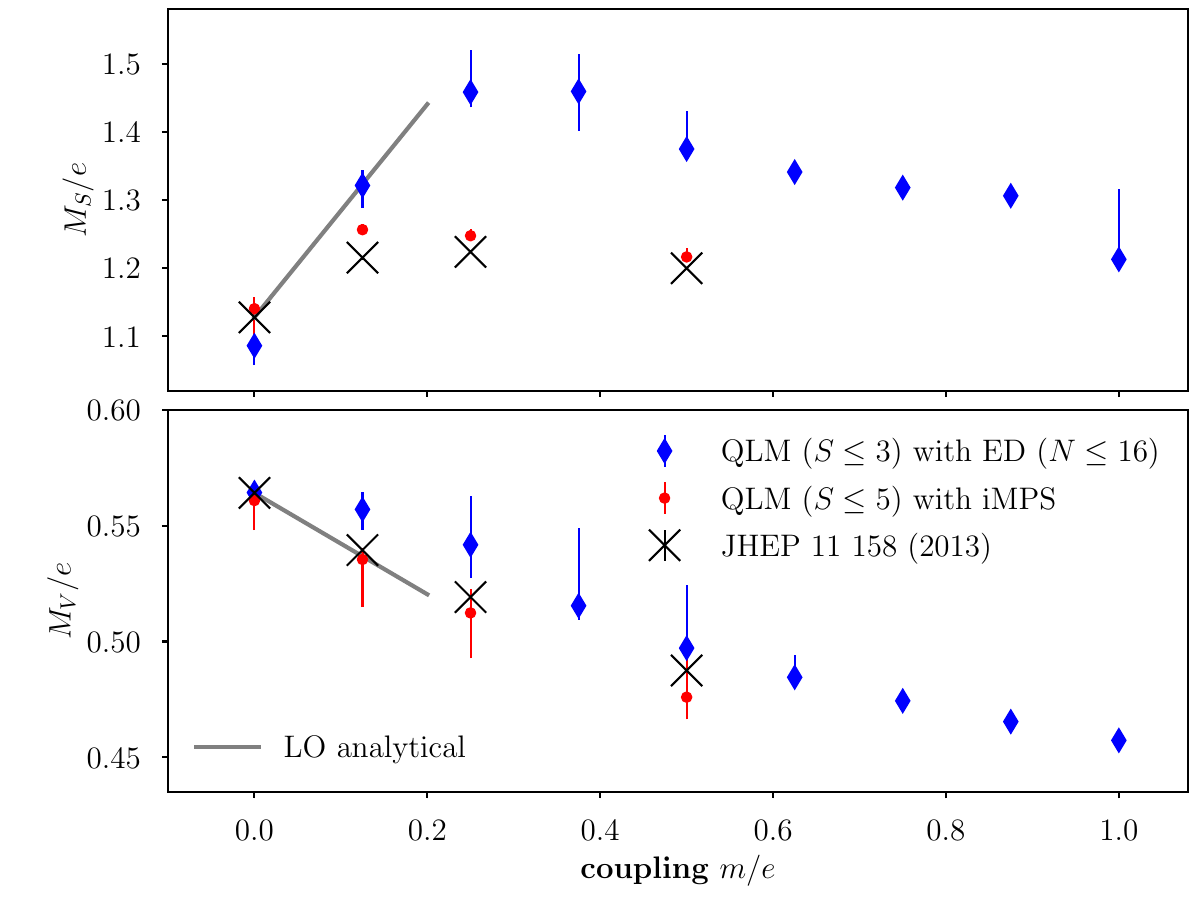}
		\caption{Final result for vector (lower panel) and scalar (upper panel) masses, with the leading dependence $2m/e$ subtracted. Our results reproduce the analytic prediction of the massless limit ($m/e=0$) and are quantitatively consistent with the perturbative expectation (grey solid line). For the vector mass, both iMPS (red dots) and ED (blue diamonds) results agree with each other and previously obtained results (black crosses) at infinite spin length.}
		\label{fig:masses}
	\end{figure}

	Figure~\ref{fig:masses} shows our final results for the vector and scalar masses with a trivial $m/e$ dependence subtracted. The iMPS results are obtained analogously to the ED simulations, but without the $N$ extrapolation. The agreement between both approaches demonstrates that  the thermodynamic limit is reached, and that the limits $S\rightarrow \infty$ and $N \rightarrow \infty$ commute for this model. Comparing to previously obtained results~\cite{banuls2013mass} at $S\rightarrow \infty$, we find good agreement of both the ED and iMPS data for the vector mass, indicating that $S=3$ is sufficient to resolve this excitation. As anticipated, the scalar mass requires larger $S$ values, and we observe stronger deviations in ED.
	
	\paragraph{Estimation of required resources on a quantum device --}
	As illustrated above, already very small spin lengths $S \lesssim 3$ and system sizes $N \lesssim 16$ are sufficient to obtain quantitative estimates for the low-lying mass spectrum. A brute-force implementation would nevertheless still require controlling a Hilbert space of dimension $\left[2(2S+1)\right]^N \sim 14^{16} \sim 2^{61}$. Due to Gauss' law, most of these states are unnecessary.
	
	To estimate the minimal required resources to implement the model on a quantum device naturally working with the corresponding qudits of size $2S+1$~\cite{kasper2020universal,Wang2020,Kiktenko_2015,Moro2019,osti_1558470,Wang_2018,Mischuck2012,Kiktenko2020}, consider the equivalent spin model, Eq.~\eqref{eq:spin_model}. Then control over only $N \sim 16$ such qudits would be sufficient to reach the continuum limit. According to Eq.~\eqref{eq:H_dim}, a perfect encoding on a digital quantum computer would need only $\text{dim} \,\mathcal{H}_\text{phys} (S=3, N=16) = 63757 < 2^{16}$ states, enabling our procedure to be carried out on existing quantum computing devices with control over $16$ qubits. This fact has been already exploited to carry out the ED calculations on a conventional laptop computer.
	
	To illustrate the applicability of our improved encoding [Eq.~\eqref{eq:H_dim}], consider two examples: (i) the mass-spectrum discussed above using a perfect qubit-encoding, and (ii) real-time dynamics on qudit hardware. The first example may be tackled with a variational quantum eigensolver (VQE) ~\cite{cerezo2021variational,atas20212} which finds an optimal representation of an input variational ansatz for the ground or a low-lying excited state, by classically minimizing $\langle \hat{H}\rangle$. We emphasize that this does not require to actually implement $\hat{H}$, but only to measure its expectation value. It is then favourable to work in a computational basis with the perfect encoding where $\hat{S}^z$ is diagonal, such that most terms of $\hat{H}$ can be measured directly. Only $\hat{S}^x$ has to be treated separately, which we leave for future work, but we note that it will remain local because the projection $\mathcal{P}$ acts locally. Having obtained the spectrum via VQE, the masses are extracted by classical post-processing via the extrapolations discussed above.
	
	For real-time dynamics, a Trotter decomposition on a system of qudits yields~\footnote{Note that here we do not enforce the perfect encoding. The Trotter decomposition is then valid on states $|\psi\rangle$ that fulfill $\mathcal{P}|\psi\rangle = |\psi \rangle$, and since $\mathcal{P}$ commutes with $\hat{S}^z_n$, the projector only appears with the terms involving $\hat{S}^x_n$.}
	\begin{align}
		e^{-i\hat{H}\Delta t} &= \prod_{n=1}^N e^{-i\frac{g^2}{2} \left(\hat{S}^z_n\right)^2\Delta t} \times   \prod_{n=1}^N e^{-i2 (-1)^n\mu \hat{S}^z_n\Delta t} 
		\nonumber \\ &\times  \prod_{n=1}^N \mathcal{P} e^{-i/\sqrt{S(S+1)}\hat{S}^x_n\Delta t} \mathcal{P} + \mathcal{O}\left(\Delta t^2\right) \;.
	\end{align}
	The first two single-qudit gates can be efficiently executed in parallel. For a fixed $n$ the off-diagonal term with $\hat{S}^x_n$  involves a projector acting locally on $(n-1,n,n+1)$. These corresponding Trotter evolution can therefore be realized as a three-qudit controlled unitary acting only on the middle qudit $n$ if the triplet $(n-1,n,n+1)$ is compatible with Gauss' law. This suggests to parallelize this last term in three layers, and thus scalable to large system sizes. This approach might be used, e.g., to observe the type of quench dynamics discussed in~\cite{halimeh2021achieving}, where the required small spin lengths $S\le 4$ are within reach of exisiting technology (see, e.g.,~\cite{wang2020qudits} for a review of qudit quantum computing and~\cite{ringbauer2021universal} for an experimental realization, as well as~\cite{morvan2021qutrit,blok2021quantum,hill2021realization,kurkcuoglu2021quantum,alam2021quantum} for recent related developments using qudits.).

	\begin{figure}[!tbh]
		\includegraphics[scale=0.8]{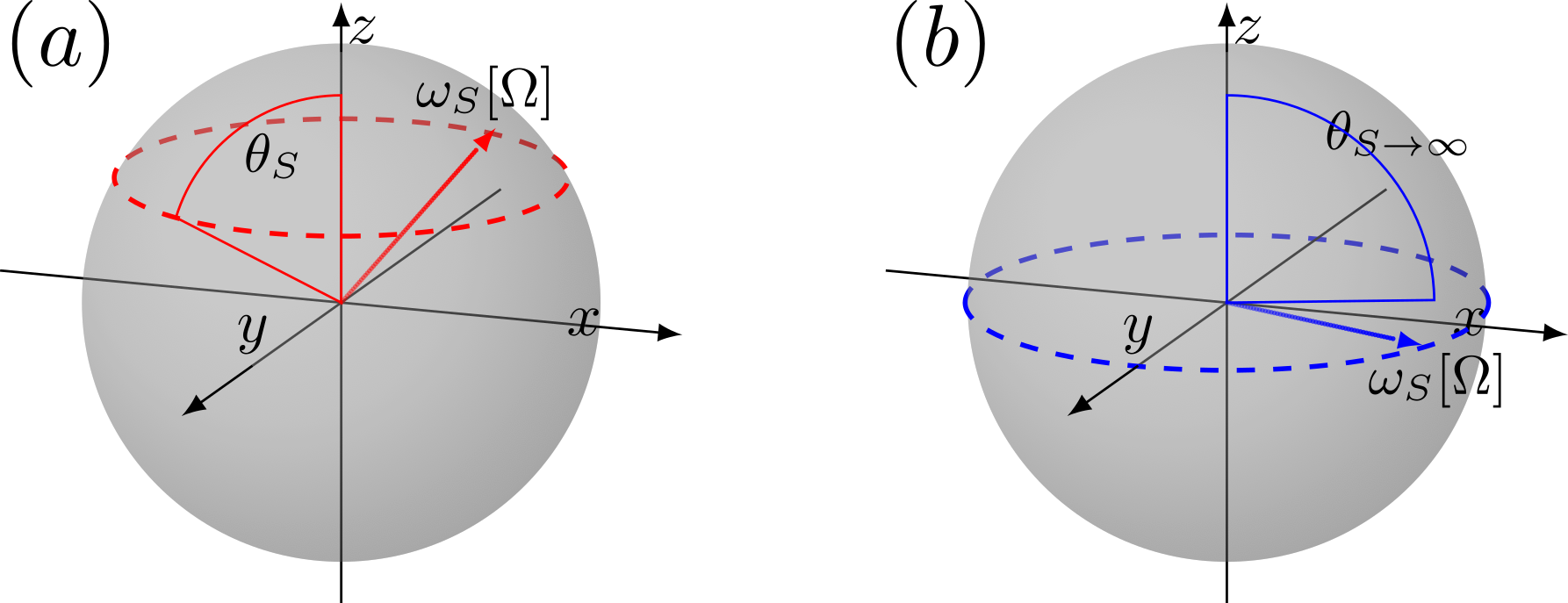}
		\caption{$\textbf{(a)}$ Closed path traced by a quantum spin-$S$ on the Bloch sphere;~$\textbf{(b)}$ with increasing $S$, the path of the spin is forced along the equator to minimize fluctuations of the electric energy term.}
		\label{fig:path}
	\end{figure}
	
	\paragraph{Field theoretic description --}
	To address the closely related question of covergence to the Kogut-Susskind limit, we use coherent state path integrals. Physically, the worldline of a spin-$S$ at spatial site ${\bf n}$ traces out an arbitrary closed curve on the Bloch sphere under the Hamiltonian evolution in imaginary time, subtending a solid angle $\Omega$, as shown in Fig.~\ref{fig:path}(a). The path integral ${\cal{Z}} = \text{Tr}_{\rm \hat{G}} \left( \text{e}^{-\beta \hat{H}}\right)$, (see Sec V of the SM) is dominated by the electric field term, $\big(\hat{S}^z_{{\bf n}}\big)^2$, which contributes as $\sim \exp\big[ -\epsilon g^2 S (S-1/2) \cos^2(\theta_{{\bf n}}) /2\big]$, where $\epsilon$ is the Trotter discretization and $\theta_{\bf n}, \phi_{\bf n}$ are the angular coordinates of the spin. For large $g^2 S$, $\theta_{{\bf n}} \rightarrow \pi/2$ as indicated in Fig.~\ref{fig:path}(b). In this limit, the quantum spin is confined to the equator of the Bloch sphere and reducse to the quantum rotor of the Wilson-Kogut-Susskind formulation, consistent with the numerics.
	
	An important additional observation is that the area traced out by the closed curve on the Bloch sphere by ${\bf \hat{S}_n}$ admits a topological interpretation as a Berry phase. In the continuum limit, this area is $\omega [\Omega] = \int_{0}^{\beta} d\tau \dot{\phi} (\tau) \cos(\theta (\phi(\tau))) = \oint_{\phi_0}^{\phi_0} d \phi \cos (\theta (\phi))$ ~\cite{Berry1984}, where $\tau \in [0,\beta]$ is the Euclidean time. At leading order in the large-$S$ limit, only the number of total windings is relevant, with each winding contributing $2 \pi S$. While this term is irrelevant for the integer spins used in this work, for half-integer spins it gives rise to a $\pi$-flux in the system. This feature rigorously establishes that QLMs with large values of half-integer spin lead to a topological theta angle $\Theta=\pi$, often anticipated in the literature ~\cite{Banerjee2012pg, Hauke_2013, surace2020lattice, yang2020observation} already for the spin-$1/2$ case.

	\paragraph{Conclusion and outlook --}
	In this paper, we have numerically demonstrated the continuum limit of QED in $(1+1)-$d, regularized with quantum spin-$S$ operators for several physical observables. A highlight of our results are the small spin values $S\lesssim 3-5$ that suffice to reach the continuum limit. The systematic finite size scaling enables us to quantitatively estimate the resources of a quantum device to realize the continuum limit. These results lend hope that near future quantum simulation experiments with small $S$ and limited lattice size can yield valuable data that can be extrapolated to the QFT limit. Using a coherent state path integral, the approach to the Kogut-Susskind limit is derived and the connection of half-integer spins with topological angle $\pi$ is formalized.
	
	For future investigations, the feasibility of our approach may immediately be tested using today's quantum hardware. High-energy aspects of these models, such as the presence of quantum scars~\cite{ho2019periodic}, and Floquet dynamics~\cite{Mukherjee2020,Mukh2} are worth studying. In higher dimensions, an exciting challenge is to quantify the convergence properties of both Abelian and non-Abelian QLMs with increasing link representations. Finally, we note that the calculation of the mass spectrum of the Schwinger model using quantum devices is merely a stepping stone towards more complex tasks. The situation changes completely when considering, e.g., real-time dynamics where our proposed Trotterization using qudits can provide a substantial advantage of traditional qubit approaches.

	\section{Acknowledgments}
	We thank Shailesh Chandrasekharan, Robert Ott, Arnab Sen, and Uwe-Jens Wiese for useful discussions. This work was supported by the Simons Collaboration on UltraQuantum Matter, which is a grant from the Simons Foundation (651440, P.Z.). D.B.~acknowledges support by the German Research Foundation (DFG), Grant ID BA 5847/2-1. This work is part of and supported by the Interdisciplinary Center Q@TN — Quantum Science and Technologies at Trento, the DFG Collaborative Research Centre SFB 1225 (ISOQUANT), the Provincia Autonoma di Trento, and the ERC Starting Grant StrEnQTh (Project- ID 804305).
	
	\bibliography{references}
	
\clearpage

\title{Supplemental Material: Towards the continuum limit of a $(1+1)$d quantum link Schwinger model}
\date{\today}

\maketitle
\usetagform{supplementary}


\section{I. Spin length and lattice units in the QLM}
Throughout the main text, we work with units where the speed of light and the Planck constant are set to one, $c = \hbar = 1$. In standard lattice QED, all quantities are rescaled with the lattice spacing $a$ to become dimensionless. For example, the dimensionless coupling is given by $g^2 = e^2 a^{3-d}$, where $e$ is the bare value of the electric charge in $d$ spatial dimensions, and $\mu = a m$ with the bare mass $m$. Similarly, the Hamiltonian is rescaled as $\hat{H}_\text{lat} = a \hat{H}_\text{phys}$ and the fermion operators are related by $\hat{\psi}_\text{lat} = a^{d/2} \hat{\psi}_\text{phys}$, such that $\left\lbrace\left(\hat{\psi}_\text{lat}\right)_\mathbf{n}, \left(\hat{\psi}_\text{lat}\right)_\mathbf{m}^\dagger \right\rbrace = \delta_{\mathbf{n}, \mathbf{m}}$.

In the QLM, dimensionless gauge fields are further replaced by spin operators. Explicitly, the electric field $\hat{E}_\text{lat} = a^{d-1}/e \hat{E}_\text{phys}$ and the compact $U(1)$ link operator $\hat{U}_\text{lat} = e^{iae \hat{A}_\text{phys}}$ are replaced according to
\begin{align}
	\hat{E}_\text{lat} \leftrightarrow  \hat{S}^z \;, &&
	\hat{U}_\text{lat}  \leftrightarrow \frac{1}{\sqrt{S(S+1)}} \hat{S}^+ \;.
\end{align}
Here, and in the following the operators $\hat{S}^{\pm,z}$ fulfill the angular momentum algebra $[\hat{S}^z,\hat{S}^{\pm}] =\pm \hat{S}^\pm$ and $[\hat{S}^+,\hat{S}^{-}] =2\hat{S}^z$ with fixed spin length $S$.
These identifications are consistent with the canonical commutation relations in the continuum, $\left[\hat{A}_\text{phys}(\mathbf{x}), \hat{E}_\text{phys}(\mathbf{y})\right] = i \delta(\mathbf{x} - \mathbf{y})$. Here, the fields with subscript ``phys'' serve as a reminder of the corresponding (dimensional) fields in the continuum theory. As a cross-check, naively sending $a\rightarrow 0$ everywhere reproduces the desired continuum expressions. Crucially, the explicit appearance of the spin length $S$ ensures the lattice commutation relations $\left[\hat{E}_\text{lat}, \hat{U}_\text{lat}\right] = \hat{U}_\text{lat}$ and $\left[\hat{U}_\text{lat},\hat{U}^\dagger_\text{lat}\right] = 0$ in the limit $S\rightarrow \infty$. To see this, consider the operator $\hat{e^{i\varphi_S}}$ which fulfills $\left[\hat{S}^z, \hat{e^{i\varphi_S}}\right] = \hat{e^{i\varphi_S}}$ exactly~\cite{Haldane1983}, given by
\begin{align}
	\hat{e^{i\varphi_S}} &= (S + \hat{S}^z)^{-1/2} \hat{S}^+ (S-\hat{S}^z)^{-1/2} \\
	&= \frac{1}{S} \left(1 - \frac{1}{2S} + \frac{1}{8S^2}\right) \hat{S}^+
	\nonumber \\& \qquad+ \frac{1}{S^3} \left[\left(\hat{S}^z\right)^2 \hat{S}^+ - \hat{S}^+ \left(\hat{S}^z\right)^2\right] + \dots \;,
\end{align}
where the omitted terms are formally suppressed for $S \gg \hat{S}^z$. Since $\left[S(S+1)\right]^{-1/2}  = \frac{1}{S} \left(1 - \frac{1}{2S} \right) + \mathcal{O}(1/S^3)$, the normalization
given in Eq.~{\usetagform{main}\eqref{eq:LGT_Hamiltonian}\usetagform{supplementary}} captures the correct behavior for large $S$.

\section{II. Equivalent spin model and Hilbert space dimension}
We derive the effective spin model starting from Eq.~{\usetagform{main}\eqref{eq:LGT_Hamiltonian}\usetagform{supplementary}} in one dimension, i.e.
\begin{align}
	\hat{H} = &\sum_{j=1}^{N}  \left[ \frac{g^2}{2} (\hat{S}^{z}_{j})^2 + \mu (-1)^j \hat{\psi}_j^\dagger \hat{\psi}_j  \right. \\
	&+ \left. \frac{1}{2\sqrt{S(S+1)}} \left(\hat{\psi}_j^\dagger \hat{S}^+_{j} \hat{\psi}_{j+1} + \text{h.c.}\right) \right]\; .
\end{align}
Here and in the following, we  we take $N$ to be even and assume periodic boundary conditions.
The Hamiltonian is gauge invariant, which can be expressed as the vanishing commutator $\left[\hat{H}, \hat{G}_j\right] = 0$ with the Gauss' law operators
\begin{align}
	\hat{G}_j = \hat{S}^z_{j} - \hat{S}^z_{j-1} - \left[\hat{\psi}_j^\dagger \hat{\psi}_j + \frac{(-1)^j-1}{2}\right]\;.
\end{align}
Additionally, the total fermion number is conserved,  i.e. $\left[\hat{H},\hat{N}_f\right] = 0$ with the fermion number operator  \begin{align}
	\hat{N}_f = \sum_{j=1}^{N} \hat{\psi}_j^\dagger \hat{\psi}_j \;.
\end{align}

As a first step, we deal with the fermionic degrees of freedom and perform  a Jordan-Wigner transformation,
\begin{align}
	\hat{\psi}_j = \hat{\sigma}^-_{j} \prod_{k=1}^{j-1} \hat{\sigma}^z_{j} \;, 
	&& \hat{\psi}_j^\dagger \hat{\psi}_j = \frac{1}{2} \left(1 + \hat{\sigma}^z_{j}\right) \;.
\end{align}
We thus obtain the equivalent Hamiltonian
\begin{align}
	\hat{H} &= \sum_{j=1}^{N} \left[ \frac{g^2}{2} (\hat{S}^{z}_{j})^2 + \frac{\mu}{2} (-1)^j \hat{\sigma}^z_{j} \right]\\
	&- \frac{1}{2\sqrt{S(S+1)}} \sum_{j=1}^{N-1}  (-1)^{\hat{\alpha}_j}   \left(\hat{\sigma}^+_{j} \hat{S}^+_{j} \hat{\sigma}^-_{j+1} + \text{h.c.}\right)\;,\nonumber
\end{align}
where $\hat{\alpha}_j= \hat{N}_f$ for $j = N$ and zero otherwise.
The full Hilbert space $\mathcal{H} = \mathcal{H}_S \otimes \mathcal{H}_\sigma$ has dimension $\dim \mathcal{H} = (2S(S+1))^{2N}$, but the physical Hilbert space $\mathcal{H}_\text{phys} = \mathcal{H} |_{G_n=0} = \left\lbrace |\text{phys} \rangle \in \mathcal{H} \;  | \; G_n |\text{phys} \rangle = 0  \right\rbrace $ is much smaller.

Within $\mathcal{H}_\text{phys}$, the configurations of $\hat{\sigma}$ are uniquely determined by those of $\hat{S}$ and thus we can express all operators in $\mathcal{H}_\text{phys}$ in terms of $\hat{S}$. Let $\mathcal{P}$ be the operator that projects all possible spin configurations in $\mathcal{H}_S$ to the allowed ones in $\mathcal{H}_\text{phys}$.  Employing $\hat{S}^z_{j} - \hat{S}^z_{j-1} = \frac{1}{2} \left[ \hat{\sigma}^z_{j} + (-1)^j\right]$ on the physical subspace and shifting indices, we find
\begin{align}
	\hat{H} = \mathcal{P}\sum_{j=1}^{N} &\left[ \frac{g^2}{2} (\hat{S}^{z}_{j})^2 + 2 \mu (-1)^j  \hat{S}^z_{j}  \right. \nonumber \\
	&- \left.\frac{1}{\sqrt{S(S+1)}} (-1)^{\alpha'_j}\hat{S}^x_{j}  \right]\mathcal{P} \; ,
\end{align}
where $\alpha'_j = N/2$ for $j=N$ and zero otherwise. In the main text, we have dropped the phase $\alpha'$, as it becomes irrelevant in the thermodynamic limit. We emphasize the importance of the projector $\mathcal{P}$, which effectively leads to an interaction among neighboring spins and allows us to rewrite
$(1/2) \mathcal{P}\left(\hat{\sigma}^+_{j} \hat{S}^+_{j} \hat{\sigma}^-_{j+1} + \text{h.c.}\right)\mathcal{P} = \mathcal{P}\hat{S}^x_{j} \mathcal{P}$ (see also \cite{surace2020lattice} for the special case of $S=1/2$).

\begin{figure}
	{\centering
		\includegraphics[scale=0.2]{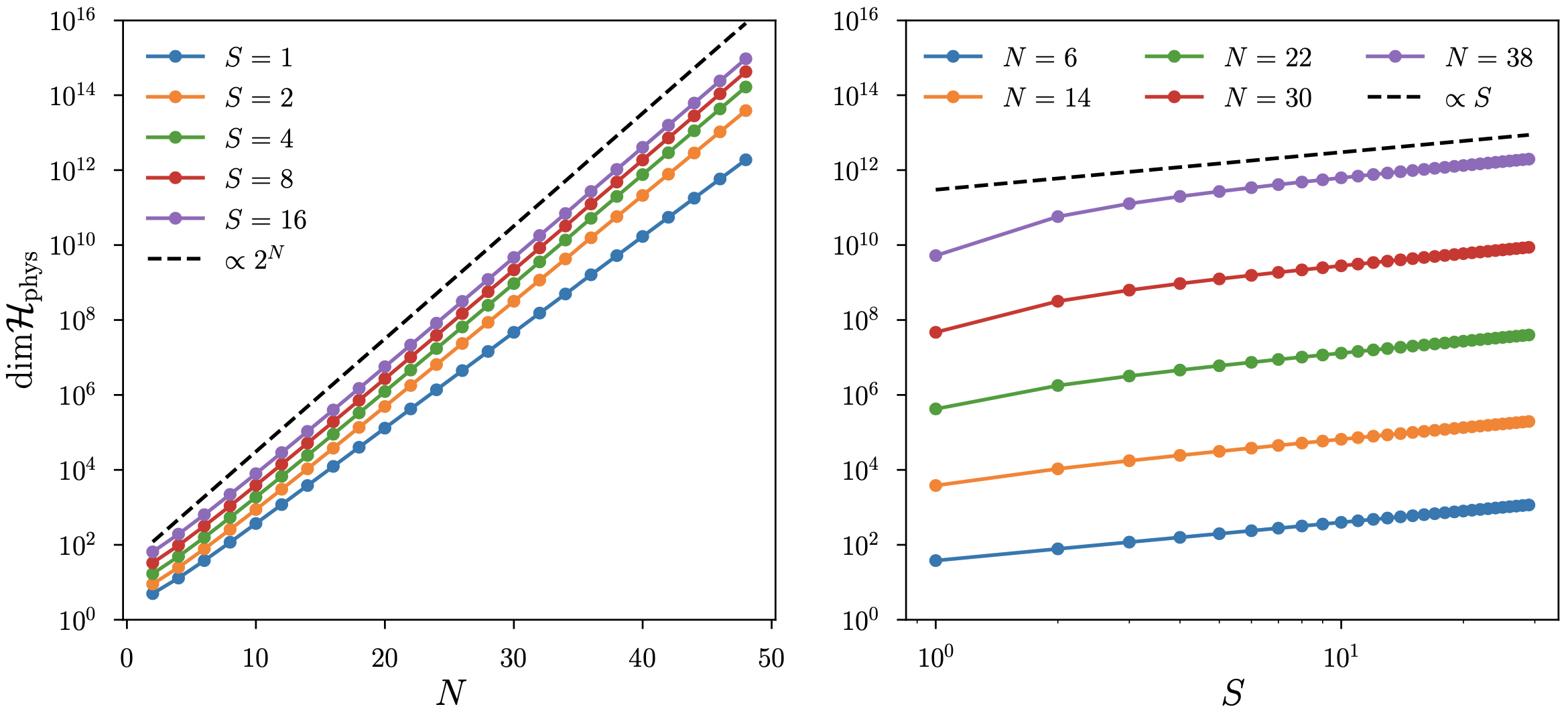}
		\caption{\label{fig:dimension_scaling} Scaling of the Hilbert space dimension [Eq.~\eqref{eq:dim_Hphys}] with system size $N$ (left) and spin length $S$ (right). For large $N$ and $S$, the dimension scales exponentially with system size, but only linearly with spin length.
	}}
\end{figure}

To find the dimension of $\mathcal{H}_\text{phys}$, we need to count all possible configurations allowed by Gauss' law. In the basis of $\hat{S}_z$, this is conveniently expressed by a square matrix $C_S$ of size $2S+1$,
\begin{align}
	C_S = \begin{pmatrix}
		1 &   \\
		1 & 1 & \\
		& 1 & 1 & \\
		& &&  \dots   \\
		&&&& 1 \\
		&&& &  1 & 1 & \\
		&&&&& 1&1
	\end{pmatrix} \; .
\end{align}
Here, rows and columns correspond to all possible spin configurations of two neighbors and $1$ or $0$ indicate an allowed or forbidden pair, respectively. Gauss' law alternates between even and odd sites as expressed by the matrices $C_S$ and $C_S^T$. The total Hilbert space dimension can now be obtained by summing over all allowed configurations, which is (for periodic boundary conditions) equivalent to the following trace
\begin{align}
	\text{dim} \mathcal{H}_\text{phys} = \text{tr} \,\left[ \left(C_S C_S^T\right)^{N/2}\right] = \sum_j \left(x^{(S)}_j\right)^{N/2} \; .
\end{align}
The eigenvalues $x^{(S)}_j$ of the matrix
\begin{align}
	C_SC_S^T = \begin{pmatrix}
		1 & 1 & \\
		1 & 2 & 1 & \\
		& 1  & 2 &\\
		&& & \dots \\
		&& && 2 & 1\\
		&&&& 1 & 2 & 1 \\
		&&&& & 1 & 2
	\end{pmatrix}
\end{align}
are given by
\begin{align}
	x^{(S)}_j = 2 + 2 \cos \left(\frac{2j \pi}{4S + 3}\right) \;, && j= 1, \dots, 2S+1 \; .
\end{align}
In summary, we proved
\begin{align}\label{eq:dim_Hphys}
	\text{dim} \mathcal{H}_\text{phys} &= \sum_{j=1}^{2S+1} \left[2 + 2 \cos \left(\frac{2j \pi}{4S + 3}\right)\right]^{N/2} \nonumber \\
	&= 2^{N}\sum_{j=1}^{2S+1} \left[\cos \left(\frac{j \pi}{4S + 3}\right)\right]^{N} \; .
\end{align}
In our main text and our exact diagonalization numerics, we make extensive use of the smallness of this dimension.  The scaling of the dimension with $S$ and $N$ is illustrated in figure~\ref{fig:dimension_scaling}.

\section{III. Analytical expressions for the vector and scalar masses}
The exact masses in the strong coupling limit ($m/e = \mu/g = 0$) are known to be
\begin{align}
	\frac{M_V^{(0)}}{e} = \frac{1}{2}\frac{M_S^{(0)} }{e}= \frac{1}{\pi} \;,
\end{align}
For the readers' convenience, we also quote the perturbative predictions which we used for the plots in the main text. We refer to~\cite{sriganesh2000new} and references therein for details.
Explicitly, the strong coupling expansions are given by
\begin{align}
	\frac{M_V+2m}{e}  &= 0.5642 + 1.781  \left(\frac{m}{e} \right) + 0.1907  \left(\frac{m}{e} \right)^2 +\dots \;,\\
	\frac{M_S+2m}{e} &= 1.128 + 3.562   \left(\frac{m}{e} \right) - 13.512  \left(\frac{m}{e} \right)^2 +\dots \;.
\end{align}

\section{IV. Extrapolations}
As described in the main text, the ground state energy, vector, and scalar mass of the continuum Schwinger model are obtained by a series of extrapolations.
From the ED, we numerically obtain the lowest three ($j=0,1,2$) eigenvalues $E_{j} = E_{j}(S,N,g,m/e)$ in the zero-momentum sector of the Hamiltonian, grouped by several values of spin length $S$, lattice size $N$, coupling $g=ae$, and mass $m/e$. In a first step, we rescale all quantities according to their expected behavior in the continuum limit (see e.g.~\cite{banuls2013mass}),
\begin{align}
	\omega_0(S,N,g,m/e) &= \frac{g^2}{2N}E_0(S,N,g,m/e) \;, \\ \omega_{1,2}(S,N,g,m/e) &= \frac{g}{2}(E_{1,2}-E_0)(S,N,g,m/e) \;.
\end{align}
Below, we indicate the step-wise extrapolation in each of the parameters $S$, $N$ and $g$ for different values of $m/e$. The extrapolated parameter is dropped in the list of arguments of the extrapolated function. For example, the extrapolations to large $S$ are obtained by fitting
\begin{align} \label{eq:fit1}
	\omega_j(S,N,g,m/e) = \omega_j(N,g,m/e) + \frac{\alpha_j(N,g,m/e)}{S} \;.
\end{align}
and the extrapolated function $\omega_j (N,g,m/e)$ does not have the $S$-dependence any more. The $1/S$ dependence in this extrapolation is motivated as the simplest polynomial dependence on $S$, and also justified a-posteriori from the data. An exponential dependence would need a scale, and we have no reason to postulate the emergence of an additional lengthscale. Similarly, the large-$N$ limit is obtained by the fits
\begin{align}  \label{eq:fit2a}
	\omega_0(N,g,m/e) &= \omega_0(g,m/e)  + \frac{\beta_0(g,m/e)}{N} + \frac{\gamma_0(g,m/e)}{N^2} \;, \\  \label{eq:fit2b}
	\omega_{1,2}(N,g,m/e) &= \omega_{1,2}(g,m/e)  +  \frac{\gamma_{1,2}(g,m/e)}{N^2} \;,
\end{align}
where we take into account that the leading finite size corrections of the excited states arise at second order in $N$~\cite{banuls2013mass}). Finally, we extrapolate to the continuum limit via
\begin{align}  \label{eq:fit3a}
	\omega_0(g,m/e) &= \omega_0(m/e)  + \delta_0(m/e) g + \epsilon_0(m/e) g^2\;,\\  \label{eq:fit3b}
	\omega_{1,2}(g,m/e) &= \omega_{1,2}(m/e)  + \delta_{1,2}(m/e) g \;,
\end{align}
where including the second order for the ground state significantly improves the accuracy of our results, indicating that our lattice data is rather far away from the continuum limit. Our final results are the ground state energy density $\omega_0(m/e)$ and the first two energies $\omega_{1,2}(m/e)$ above the ground state.
The extrapolations of the iMPS data proceed analogously, omitting the fit with respect to $N$.

In general, the data $\omega_j(S,N,g,m/e)$ diverges as $g \rightarrow 0$ for fixed $S,N$, and $m/e$ in such a way that for every $g$ there are minimal values of $S$ and $N$ required for a reliable extrapolation. In order to avoid additional assumptions about the functional form of $\omega_j(S,N,g,m/e)$, we have implemented a simple convergence check to discard data with too small values of $S$ and $N$: To obtain the results presented in the main text, we discarded all bare data corresponding to the same value of $g$ (and fixed $j$) if the extrapolated values $\omega_j(N,g,m/e)$ differ by more than $10\%$ from the value corresponding to the largest available spin length $S_\text{max}$, i.e. when $|\omega_j(N,g,m/e) - \omega_j(S_\text{max},N,g,m/e)| > 0.1 \times  |\omega_j(S_\text{max},N,g,m/e)|$. For the ED data, we checked for convergence of the $N$ extrapolation analogously. After supplementing the fits by this procedure, and given the relatively small values of $S$ (and $N$ for the ED), we find that the uncertainty of all quantities is dominated by the final extrapolation to the continuum limit. The error bars shown in the plots of the main text indicate systematic errors that arise from the corresponding choice of fit range. We have estimated these errors following the procedure described in the appendix of~\cite{banuls2013mass}.  Our final results are summarized in figure~\ref{table:results}, { where we also state the normalized square error 
	\begin{align}
		\chi^2 = \frac{1}{N_{\text{data}}}\sum_{k=1}^{N_{\text{data}}} \left[ y_k - f(y_k) \right]^2  \;,
	\end{align}
	to illustrate that the errors of individual fits are negligible.
	Here $y_k$ denotes the data used for the fit [e.g.~$\omega_k(S, \dots)$ for several spin lengths $S$], $f(y_k)$ is the fit result evaluated at $y_k$ [with $f$ the appropriate fit function from Eq.~\eqref{eq:fit1},~\eqref{eq:fit2a},~\eqref{eq:fit2b},~\eqref{eq:fit3a} or ~\eqref{eq:fit3b}] and $N_\text{data}$ is the number of data points $y_k$ used for the particular fit. We emphasize that the $\chi^2$-value has no statistical interpretation as all our calculations are purely deterministic.
} 

\begin{figure}
	\centering
	\begin{tabular}{c || c | c | c}
		$m/e$ & $\omega_0$ & $M_V/e$ & $M_S/e$\\
		\hline \hline
		0 & $-0.3205_{-0.0045}^{+0.0037}$& $0.5643_{-0.0046}^{+0.0010}$ & $1.086_{-0.028}^{+0.038}$ \\
		\hline
		0.125 & $-0.3204_{-0.0053}^{+0.0036}$ & $0.5571_{-0.0088}^{+0.0076}$ & $1.321_{-0.033}^{ +0.023}$\\
		\hline
		0.25 & $-0.3198_{-0.0042}^{+0.0030}$ & $0.5418_{-0.0145}^{+0.0213}$ & $1.458_{-0.022}^{+0.062}$ \\
		\hline
		0.375 & $-0.3188_{-0.0027}^{+0.0020}$ & $0.5155_{-0.0061}^{+0.0336}$ & $1.459_{-0.058}^{ +0.055}$ \\
		\hline
		0.5 & $-0.3187_{-0.0011}^{+0.0013}$ & $0.4972_{-0.0017}^{+0.0275}$ & $1.375_{-0.009}^{+0.056}$ \\
		\hline
		0.625 & $-0.3169_{-0.0001}^{+0.0003}$ & $0.4845_{-0.0008}^{+0.0095}$ & $1.341_{-0.006}^{+0.016}$ \\
		\hline
		0.75 & $-0.3156_{-0.0006}^{+0.0003}$ & $0.4744_{-0.0005}^{+0.0044}$ & $1.318_{-0.007}^{+0.007}$ \\
		\hline
		0.875& $-0.3147_{-0.0013}^{+0.0004}$ & $0.4654_{-0.0004}^{+0.0010}$
		& $1.306_{-0.008}^{+0.009}$ \\
		\hline
		1& $-0.3143_{-0.0014}^{+0.0006}$ & $0.4573_{-0.0005}^{+0.0003}$ & $1.213_{-0.003}^{+0.104}$  
	\end{tabular}
	\caption{\label{table:results}Final results for the ground state energy $\omega_0$ and the vector(scalar) masses $M_{V(S)}$ for the values of $m/e$ that correspond to the data shown in the main text.}
\end{figure}

\section{V. Path integral representation}
In this section, we outline a derivation of the path integral for the pure gauge $U(1)$ QLM Hamiltonian using coherent states for the quantum spin operators representing the gauge fields, and indicate how the large representations give rise to the Wilson-Kogut-Susskind limit of the gauge theory. Given a Hamiltonian $H$ in $d-$spatial dimensions which satisfies a local constraint, the corresponding path integral is given by 
\begin{equation}
	{\cal{Z}} = \text{Tr} \left( \text{e}^{-\beta H}\mathbb{P} \right) 
	= \int \mathcal{D} \Omega ~ \text{e}^{-S [\Omega]}.
\end{equation}
The extent in Euclidean time, $\beta$ equals inverse temperature T, and $\mathbb{P}$ is the projection operator projecting the configurations in a chosen computational basis to the ones allowed by the local constraint. The path integral is constructed by splitting the total Euclidean time $\beta$ into $N_t$ Trotter steps of extent $\epsilon$, such that $\beta = \epsilon N_t$. At each Trotter step, we construct the transfer matrix, $\Braket{\Omega_{\rm n-1}| \mathbb{T} | \Omega_{\rm n} } = \Braket{\Omega_{\rm n-1}| \text{exp} (-\epsilon H) | \Omega_{\rm n} } $, where $\ket{\Omega_{\rm n}}$ denotes the computational basis (which we introduce next) at time-slices $\text{n-1}$ and $\text{n}$. A periodic boundary condition in the imaginary time direction for the gauge field is used. The final expression is an integral over all possible field configurations $\Omega$ in $(d+1)-$dimensions in the limit $\epsilon \to 0, N_t \to \infty$ with $\beta$ held fixed. Further, for $\beta \to \infty$ ground state results can be obtained.

The pure gauge Hamiltonian is given by $H_{\text g} = \frac{g^2}{2} \sum_{j=1}^N (\hat{S}^z_j)^2$, and the Gauss' law without the matter fields is simply $\hat{G}_j = \hat{S}^{z}_j - \hat{S}^{z}_{j-1}$, and in the following we will denote the projection operator which selects configurations according to this (local) constraint as $\mathbb{P}_G = \prod_j \mathbb{P}_{\hat{G}_j}$. We emphasize that the fermions are not included in this Hamiltonian, and consequently the Gauss' law differs from the one in the main text. To derive the path integral, we use the coherent state basis (also known as Bloch states in the literature) as the computational basis, which is denoted as ($j$ denotes a spatial site, and ${\rm n}$ the timeslice)
\begin{equation}
	\begin{split}
		\ket{\Omega_{\rm n}} &= \prod_{j} \ket{\theta_{j,{\rm n}},\phi_{j,{\rm n}}} \\
		&= \prod_{j} (\cos \frac{\theta_{j, {\rm n}}}{2})^{2S} \exp \left[  \tan(\frac{\theta_{j,{\rm n}}}{2}) {\rm e}^{i \phi_{j, {\rm n}}} S_{j,{\rm n}}^- \right] \ket{0},
	\end{split}
\end{equation}
where $0 \leq \theta_{j,{\rm n}} < \pi$ and $0 \leq \phi_{j,{\rm n}} < 2 \pi$. In what follows, we further need the resolution of identity in the coherent state basis at each timeslice n: 
\begin{equation}
	\mathbb{I} = \prod_j \frac{2S+1}{4 \pi} \int^{\pi}_{0} \sin(\theta_{j,{\rm n}}) d \theta_{j,{\rm n}} \int^{2\pi}_{0} d\phi_{j,{\rm n}} \ket{\Omega_{\rm n}} \bra{\Omega_{\rm n}}.
\end{equation}
Using the above expressions, in the coherent state basis, we write the partition function ${\cal Z}$ as
\begin{align}
	{\cal Z} &= \int {\cal D} \Omega \Braket{\Omega_0| {\rm e}^{-\epsilon H_{\rm g}} | \Omega_1 } 
	\cdots \Braket{\Omega_{N-1}| {\rm e}^{-\epsilon H_{\rm g}} | \Omega_{N}} \nonumber \\
	&\times \Braket{\Omega_N | \mathbb{P}_G | \Omega_0 }
\end{align}
The overlap $\bra{\Omega_{\rm n-1}} \Omega_{\rm n}\rangle$ constitutes an important piece in the full calculation and gives rise to the Berry phase. After some algebra, it can be expressed as 
\begin{equation}
	\Braket{\Omega_{\rm n-1} |\Omega_{\rm n}} = {\rm exp} \left[ i S \sum_j \cos(\theta_{j,{\rm n}}) (\phi_{j,{\rm n-1}} - \phi_{j,{\rm n}})  \right],
\end{equation}
where we have assumed that $\epsilon$ is small such that the fields on adjacent time-slices are sufficiently smooth. 
The matrix elements of the following operators in the coherent state basis are also needed:
\begin{equation}
	\begin{split}
		\Braket{\Omega_{\rm n-1} | \sum_j \hat{S}^z_{j,{\rm n}} | \Omega_{\rm n} } &= \Braket{\Omega_{\rm n-1} | \Omega_{\rm n}} S \sum_j \cos (\theta_{j,{\rm n}}) \\
		\Braket{\Omega_{\rm n-1} | \sum_j (\hat{S}_{j,{\rm n}}^z)^2 | \Omega_{\rm n} } &= \Braket{\Omega_{\rm n-1}|\Omega_{\rm n}} \\ 
		&\times \left[ S (S-\frac{1}{2}) \sum_j \cos^2 (\theta_{j,{\rm n}}) + \frac{S}{2} \right].
	\end{split}
\end{equation}
Combining the above formulae, the transfer matrix between two time-slices is thus
\begin{multline}
	\Braket{\Omega_{\rm n-1}| {\rm exp} (-\epsilon H_{\rm g}) | \Omega_{\rm n} } \\
	\approx \Braket{\Omega_{\rm n-1}| \Omega_{\rm n} }
	{\rm e}^{-\frac{\epsilon g^2}{2} \sum_{j=1}^N S (S - \frac{1}{2}) \cos^2(\theta_{j, {\rm n}}) - \frac{\epsilon g^2 S N}{2}},
	\label{eq:tmat}
\end{multline}
where we have expanded the exponential, evaluated the operator, and re-exponentiated it, valid up to corrections at $O(\epsilon^2)$. The last timeslice contains the projection operator, and can be written in the $\ket{\Omega}$-basis as:
\begin{gather}
	\Braket{\Omega_N | \prod_j \mathbb{P}_{\hat{G}_j} | \Omega_0 } 
	= \Braket{\Omega_N| \prod_j \delta (\hat{S}^z_j - \hat{S}^z_{j-1} ) | \Omega_0 } \nonumber \\
	= \Braket{\Omega_N |\Omega_0} \prod_j \int_{-\pi}^{\pi} \frac{d \varphi_{j}}{2 \pi} 
	\exp \left[ i \varphi_{j} S (\cos \theta_{j,{\rm 0}} - \cos \theta_{j-1,{\rm 0}})  \right] \nonumber \\
	\hspace{-1.5cm}
	= \prod_j \int_{-\pi}^{\pi} \frac{d \varphi_{j}}{2 \pi} 
	\exp \left[ i S \cos\theta_{j,0} (\phi_{j,{\rm N}} - \phi_{j,0}) \right] \nonumber \\
	\hspace{-1.5cm} \times \exp \left[ i \varphi_{j} S (\cos \theta_{j,{\rm 0}} - \cos \theta_{j-1,{\rm 0}}) \right] \;.
\end{gather}
The $\varphi_j$ are the Polyakov loop phases which appear as the Lagrange multiplier enforcing Gauss' law on the last time-slice. To incorporate the Polyakov loop variables with the action, we consider the matrix elements of the Hamiltonian and the Gauss' law together on the last two timeslices:
\begin{multline}
	\int \frac{d \Omega_{\rm N}}{4 \pi} \Braket{\Omega_{N-1}| {\rm e}^{-\epsilon H_{\rm g}} | \Omega_{\rm N}} \Braket{\Omega_{\rm N}| \prod_j \mathbb{P}_{\hat{G}_j} | \Omega_0} \\
	\hspace{-3cm}
	= \prod_j \frac{2S+1}{4 \pi} \int_0^\pi \sin \theta_{j,{\rm N}} d\theta_{j,{\rm N}} \int_{-\pi}^{\pi} d \phi_{j,{\rm N}} \nonumber \\
	\times {\rm e}^{\left[ -\frac{\epsilon g^2}{2} S (S-\frac{1}{2}) \cos^2 \theta_{j,{\rm N}} - \frac{\epsilon g^2 S N}{4} + i S \cos \theta_{j,{\rm N}} (\phi_{j, {\rm N-1}} - \phi_{j, {\rm N}})\right] } \nonumber \\
	\times \prod_j \int_{-\pi}^{\pi} \frac{d \varphi_j}{2 \pi} {\rm e}^{\left[ i \varphi_j S (\cos \theta_{j, 0} - \cos \theta_{j-1, 0}) + i S \cos \theta_{j,0} (\phi_{j,{\rm N}} - \phi_{j, 0}) \right]} \;.
\end{multline}
We collect the terms in $\phi_{j,{\rm N}}$ and perform the integral
\begin{multline}
	\prod_j \int_{-\pi}^{\pi} d\phi_{j,{\rm N}} \hspace{0.2cm} {\rm e}^{i S (-\cos \theta_{j,{\rm N}} + \cos \theta_{j, 0}) \phi_{j,{\rm N}}} \nonumber 
\end{multline}
which enforces the condition $\theta_{j, 0} = \theta_{j, {\rm N}}$ for all the spatial lattice sites (recall that $0 \leq \theta_{j,{\rm n}} < \pi$). We can use this to simplify the matrix element
\begin{multline}
	\int \frac{d \Omega_{\rm N}}{4 \pi} \Braket{\Omega_{N-1}| {\rm e}^{-\epsilon H_{\rm g}} | \Omega_{\rm N}} \Braket{\Omega_{\rm N}| \prod_j \mathbb{P}_{\hat{G}_j} | \Omega_0} \\
	\hspace{-2.5cm}
	= \prod_j \frac{2S+1}{4 \pi} \int_0^\pi \sin \theta_{j,{\rm N}} d\theta_{j,{\rm N}} \int_{-\pi}^{\pi} \frac{d \varphi_j}{2 \pi} \\
	\hspace{-3cm}
	\times {\rm e}^{\left[ -\frac{\epsilon g^2}{2} S (S-\frac{1}{2}) \cos^2 \theta_{j,{\rm N}} - \frac{\epsilon g^2 S N}{4} \right]} \\
	\hspace{-3cm}
	\times {\rm e}^{\left[i S \cos \theta_{j,0} (\phi_{j,{\rm N-1}} - \phi_{j,0} + \varphi_j - \varphi_{j+1}) \right]} \;. \nonumber
\end{multline}

In the limit $S \to \infty$, the effective action can be obtained by minimizing the quadratic term in $\cos \theta_{j,{\rm n}}$. This gives $\theta_{j, {\rm n}} \approx \frac{\pi}{2} \pm \alpha_{j,{\rm n}}$, thus $\cos \theta_{j,{\rm n}} \approx \cos (\frac{\pi}{2} - \alpha_{j,{\rm n}}) = \sin \alpha_{j,{\rm n}} \approx \alpha_{j,{\rm n}}$, and $\sin \theta_{j,{\rm n}} \approx 1$. The integral then becomes Gaussian, which can be performed explicitly. Further, we note that the term $\phi_{j,{\rm N-1}} - \phi_{j,0} = 2 \pi n_j + \Delta_t \phi_j$, where $\Delta_t \phi_j$ is the difference in the $\phi$ variable across the last timeslice. Note that the the same considerations as discussed here also hold for the other timeslices, as explicitly shown in Eq.~{\usetagform{main}\eqref{eq:tmat}\usetagform{supplementary}}. In fact, it is even simpler there, since the Gauss' Law does not need to addressed, and again the quadratic term can be extremized to obtain a Gaussian integral. The overlap of the coherent states, contributing to the Berry phase will be addressed separately.

As the analysis reveals, in the $S \to \infty$ limit, the unit vector is forced to point at the equator, with minor fluctuations (denoted by $\alpha_{j,{\rm n}}$). At every time slice overlap, the $\phi_{j,{\rm n}}$ proceeds on the unit sphere, and thus the term $\phi_{j,{\rm N-1}} - \phi_{j,0}$ contains the total number of windings around the equator. In the large-$S$ limit, we get
\begin{multline}
	\int \frac{d \Omega_N}{4 \pi} \Braket{\Omega_{N-1}| {\rm e}^{-\epsilon H_{\rm g}} | \Omega_N} \Braket{\Omega_N| \prod_j \mathbb{P}_{\hat{G}_j} | \Omega_0} \\
	\approx \sum_{n_j \in \mathbb{Z}} \sqrt{\frac{2 \pi}{S (2S-1) \epsilon g^2}} 
	{\rm e}^{\left[-\frac{S}{2(2S-1)\epsilon g^2} (\varphi_j - \varphi_{j+1} + 2\pi n_j + \Delta_t \phi_j)^2  \right]}
\end{multline}
The term in the exponential is the well-known Villain form~\cite{Villain1974} of the plaquette action in the Euclidean theory. Identifying $\varphi_{j} - \varphi_{j+1} \sim -\partial_j A_0(0)$ and $\Delta_t \phi_{j,0} \sim \partial_t A_1(0)$, the term in the exponential is the Maxwell action, and can be interpreted as the path integral in the axial gauge. In the axial gauge, all the time-like links are set to unity, except for the ones which cross the temporal boundary. In order to bring the expression to a form which is explicitly space-time symmetric, we can perform gauge transformations to spread the $A_0$ field away from the boundary time-slice and into the whole space-time lattice. The Boltzmann weight of the effective action contributing to ${\cal Z}$ is thus
\begin{equation}
	{\rm e}^{-{\cal S}_{\rm eff}} =  {\rm e}^{- \frac{1}{2 \epsilon g^2} \int dx dt (\partial_t A_1 (x,t) - \partial_x A_0 (x,t) )^2},
\end{equation}
where we have traded $n$ and $j$ for $t$ and $x$ respectively, and $A_{0,1}$ is the vector potential for the gauge field in 2-dimensions. We have thus demonstrated that how in the limit of large-$S$ the localization of the spin operator along the equator is responsible for the emergence of the Wilson lattice gauge theory in Euclidean space.

Next, we connect the contribution from the Berry phase term of the quantum spin operators to the presence of topological terms in the Wilson action. We note, following Coleman~\cite{Coleman:1976uz}, that the Hamiltonian of the continuum massless Schwinger model in the axial gauge ($A_0=0$) can be written as 
\begin{equation}
	H = \int dx \left[ -i \bar{\psi} \gamma^1 (\partial_1 + i g A_1) \psi + \frac{1}{2} E^2 \right],
\end{equation}
where $\psi, \bar{\psi}$ are the fermion fields and $\gamma^{1,2}$ are the gamma matrices. In one dimension, due to the Gauss' law, the  $E$ is not an independent operator, but can be written in terms of the charge density up to an integration constant. It was shown by Coleman~\cite{Coleman:1976uz} that the integration constant acts as a background field, has the characteristics of a periodic function and is naturally related to the topological angle. Coleman further showed that for certain special values of this topological angle, $\Theta = \pi$, the physics of the model is very different from the expected confining behavior. 

In the context of QLMs, it is quite remarkable that the use of full or half-integer spin representations can induce qualitatively different behavior, at least in certain parameter regimes~\cite{Banerjee2012pg}. In particular, using the toolbox of coherent states, we can demonstrate that for large-$S$, half-integer spin representations can give rise to the topological angle $\Theta = \pi$. 

Note that this topological term is generated from the overlap of the coherent states $\braket{\Omega_0 | \Omega_1} \braket{\Omega_1 | \Omega_2} \cdots \braket{\Omega_{N-2} | \Omega_{N-1}}= \exp(i S \sum_{j} \omega_j [\Omega_j])$ (see Eq.~{\usetagform{main}\eqref{eq:tmat}\usetagform{supplementary}}), which is the Berry phase in the problem. $\omega_j [\Omega_j]$ is the solid angle corresponding to the worldline (closed curve due to periodic boundary condition) of a spin at site-j as it evolves in imaginary time. While the Maxwell term, ${\cal S}_{\rm eff}$ is independent of $S$ in the $S \to \infty$ limit, this term is sensitive to the integer vs.~half-integer nature of the spin $S$. To see this, note that Berry phase term can be written as~\cite{Berry1984} (we drop the spatial index without loss of generality)
\begin{align}
	\omega [\Omega] = \int_0^\beta d\tau \dot{\phi} \cos(\theta_\phi) = \oint_{\phi_0}^{\phi_0} d \phi \cos (\theta_\phi).
\end{align}
Here, $\theta$ is treated as a function of $\phi$, as the spin S traces out a path $\vec{S}(\tau)$ on the surface of the 2-sphere, $S^2$. Due to the periodic boundary conditions in time, the angle $\phi_0$ must come back to itself. As we argued before, in the large-$S$ limit, the unit vector is restricted to the equator with fluctuations, which makes the phase space effectively a circle. Due to the topological nature of this term, it only matters how many windings the vector makes around the equator. Thus in the leading order, the topological term contributes as $2 \pi  S$. For half-integer $S$, this gives an effective phase $\pi$, which contributes the same way as the background field $\Theta = \pi$ in the Wilson lattice gauge theory. On the other hand, for integer values of $S$, the phase of $2 \pi$ is irrelevant, and only the plaquette action remains.

Thus, we have shown that in the large-$S$ limit, the half integer values of $S$ produce the same effective action as that of the Wilson theory with a background field of $\pi$, while the integer values produce an overall phase of multiples of $2\pi$, and no extra effects. This derivation can be straightforwardly repeated with the inclusion of fermions.


\end{document}